\documentstyle[aps,epsfig,prl,multicol]{revtex}

\relax   
\begin{document}
\preprint{prl}
\draft

\title{Cyclotron Resonance in the Layered Perovskite Superconductor Sr$_2$RuO$_4$}

\author{S. Hill\cite{email}}
\address{Department of Physics, Montana State University, Bozeman, MT 59717}

\author{J. S. Brooks}
\address{Department of Physics and National High Magnetic Field Laboratory, Florida State University,
Tallahassee, FL 32310}

\author{Z. Q. Mao and Y. Maeno}
\address{Department of Physics, University of Kyoto, Kyoto 606-8502,
and CREST-JST, Kawaguchi, Saitama 332-0012, Japan}

\date{\today}
\maketitle

\begin{abstract}
We have observed cyclotron resonance in the layered perovskite
superconductor Sr$_2$RuO$_4$. We obtain cyclotron masses for the
$\alpha$, $\beta$ and $\gamma$ Fermi surfaces of
(4.33$\pm$0.05)$m_e$, (5.81$\pm$0.05)$m_e$ and (9.71$\pm$0.2)$m_e$
respectively. The appreciable differences between these results
and those obtained from de Haas- van Alphen measurements are
attributable to strong electron-electron interactions in this
system. Our findings appear to be consistent with predictions for
an interacting Fermi liquid; indeed, semi-quantitative agreement
is obtained for the electron pockets $\beta$ and $\gamma$.
\end{abstract}


\begin{multicols}{2}[]

The perovskite superconductor Sr$_2$RuO$_4$ is currently the
subject of intense research activity \cite{Maeno1}. Initial
interest was driven by the structural similarities between this
compound and the high-$T_c$ cuprates \cite{Maeno2}. However, a
clear picture has since emerged in which it apparently belongs to
an entirely different class of superconductor. In particular,
Sr$_2$RuO$_4$ shares many properties with liquid $^3$He. For
example: the normal state conforms to the behavior expected in a
strongly correlated Fermi liquid (FL) \cite{Mack1,Maeno3}; the
highest reported $T_c$ is still relatively low ($<$ 1.5 K); $T_c$
is extremely sensitive to non-magnetic impurities \cite{Mack2};
and there is mounting evidence supporting the view that the
superconductivity is induced by ferromagnetic fluctuations
specific to quasi-two-dimensionality, resulting in spin-triplet
pairing \cite{Rice,Ishida}.

One of the many attractive features of the title compound has been
the availability of high quality single crystals. This has enabled
experimental investigations which have not been possible in the
high-$T_c$ counterparts, including de Haas - van Alphen (dHvA) and
Shubnikov - de Haas (SdH) measurements \cite{Mack1,Mack3}.
Consequently, an extensive body of experimental data has given
rise to an increasingly coherent picture of the normal state
electronic structure \cite{Mack3} which is in broad agreement with
the calculated band structure obtained by Local Density
Approximations (LDA) \cite{LDA}. However, as with many strongly
correlated electron systems, there is a considerable discrepancy
between the calculated and measured density of states at the Fermi
energy ($E_F$); the band masses ($m_b$) for the three Fermi
surface (FS) pockets (two electron and one hole), estimated from
LDA calculations, are considerably smaller (by factors of between
3 and 5) than the thermodynamic masses ($m^*$) determined from
dHvA measurements \cite{Mack1}. These differences have been
attributed to the FL corrections expected in an interacting Fermi
system.

The aim of the present study is to measure the cyclotron masses
($m_c$) in Sr$_2$RuO$_4$ using a long wavelength probe ({\em
Q}$\sim$0) which couples to the center-of-mass motion of the
system. For a translationally invariant isotropic FL, the band
mass, cyclotron mass and thermodynamic mass represent different
physical quantities \cite{Quader,kanki}. The thermodynamic mass
includes enhancements due to the fact that, as quasiparticles move
through a medium, they experience a drag force resulting from the
displacement of other quasiparticles. Thus, in strongly
interacting Fermi systems, $m^*$ includes corrections not
ordinarily included in band calculations. A cyclotron resonance
(CR) experiment is insensitive to these FL effects due to the
absence of quasiparticle drag in the center-of-mass frame.
Therefore, a comparison between the CR mass $m_c$, and the
thermodynamic mass $m^*$ (as measured {\em e.g.} in a dHvA
experiment), offers the unique possibility of gauging the
magnitude of these FL corrections. We wish to point out, however,
that the CR mass (sometimes referred to as the dynamical mass)
{\em is} sensitive to other many-body effects, $e.g$. from phonon
and Coulomb interactions (distinct from FL effects, see ref.
\cite{Quader}). Consequently, one expects some enhancement of the
CR masses ($m_c$) relative to the band masses ($m_b$).

dHvA experiments on Sr$_2$RuO$_4$ have confirmed the existence of
three quasi-two-dimensional FSs whose cross sections are weakly
modulated ($<1.5 \%$ variation in {\em k$_F$}) along the
{\bf{$c$}}-direction \cite{Mack1,Mack3,berg}. Application of a
magnetic field causes carriers to orbit these roughly cylindrical
FSs in a plane perpendicular to the field. The dominant result is
cyclotron motion within the conducting layers, even when the field
is tilted well away from the {\bf{$c$}}-axis. It is this cyclotron
motion that one usually couples to in a CR experiment. A more
subtle effect concerns the influence of a magnetic field on the
carrier velocities parallel to the cylinder axes. Depending on the
field orientation, and on the symmetry of the warping, the
{\bf{$c$}}-axis velocities ($v_c$) will also oscillate (for a
detailed discussion, see ref \cite{Hill1}); it is precisely these
$v_c$ modulations which are responsible for the Angle-dependent
Magnetoresistance Oscillations (AMRO) observed recently in
Sr$_2$RuO$_4$ \cite{Omichi}. In principle, it should also be
possible to couple directly to these periodic $v_c$ modulations in
an AC measurement, as has recently been demonstrated for several
low-dimensional organic conductors \cite{Ardavan,Hill2}.
Cyclotron-like resonances observed by this method offer a powerful
means (over an above AMRO) of determining the precise FS
topologies of low-dimensional conductors \cite{Hill1,Blundell}.
With this in mind, we opted for an experimental configuration (see
below) which is sensitive to both the in-plane and inter-layer
conductivities, $\sigma_{ab}$ and $\sigma_c$ respectively.

Because of the large effective masses in Sr$_2$RuO$_4$, it was
necessary to conduct experiments at the lowest frequencies allowed
by the constraint $\omega\tau > 1$, and to work at high magnetic
fields. The high quality (long $\tau$) of the Sr$_2$RuO$_4$ single
crystal used in this study, which was grown by a floating zone
method [4] and has a $T_c$ of 1.44 K (mid-point), enabled
measurements in the mm-wave spectral range. Fields of up to 33
tesla were provided by the resistive magnets at the National High
Magnetic Field Laboratory in Florida.

One of the benefits of working at GHz frequencies is the
possibility of utilizing an extremely sensitive cavity
perturbation technique \cite{Hill4,dress}. Two different
cylindrical copper cavities ($\phi \sim10$mm, height $\sim10$mm)
were used in transmission, providing four TE01$n$ ($n$ = 1, 2 and
3) modes in the desired frequency range. Loaded cavity $Q$-factors
ranged from $5\times10^3$ to $2\times10^4$, depending on the mode.
A single Sr$_2$RuO$_4$ crystal (dimensions $\approx 2.5 \times 1
\times <0.2 mm^3$) was placed close to the bottom of the cavity,
half way between its axis and its perimeter, thereby ensuring
optimal coupling to the radial AC magnetic fields ({\bf\~ H}$_1$)
for a given TE01{\em n} mode. In this configuration ({\bf\~
H}$_1$//{\bf{$ab$}}-plane), the microwave fields excite both
in-plane and inter-layer currents (see inset to Fig. 1 and ref.
\cite{Hill4}). The applied DC magnetic field ({\bf B}$_o$) was
directed along the cavity axis and, therefore, parallel to the
sample $c$-axis, {\em i.e.} {\bf\~ H}$_1\bot${\bf B}$_o$. As a
spectrometer, we used a Millimeter-wave Vector Network Analyzer
(MVNA) \cite{Hill4}. Finally, the cavity containing the sample
could accurately and controllably be maintained at any temperature
between 1.4 K and 30 K.



Fig. 1a shows changes in absorption within the cavity as a
function of magnetic field for several temperatures in the range
1.4 to 6 K. The cavity was excited at 76.4 GHz, which corresponds
to its TE013 mode. The data were obtained after subtracting a
background cavity response, and have been offset for the sake of
clarity. It is apparent that on cooling below $\sim$5 K, a series
of absorption peaks develop (indicated by arrows) and grow
stronger. To within the confidence of the Lorentzian fits to the
data, the peak positions appear to be independent of temperature.
The number of Lorentzians (5 at T=1.4K), and the initial estimates
of their peak positions, were chosen after inspecting data
obtained at four different frequencies (see Fig. 3) $-$ otherwise,
all fitting parameters were free running. Fig. 1b shows an almost
identical data set obtained at the lower frequency of 64.0 GHz,
which corresponds to the TE013 mode of the second cavity. It
should be noted that all of the absorption peaks have shifted to
lower magnetic field.

Similar data were obtained at 44.5 GHz and 58.5 GHz, corresponding
to the TE011 and TE012 modes of the first cavity respectively.
However, the absorption peaks were less clearly resolved from each
other at these lower frequencies due to the lower $\omega\tau$
product. In the inset to Fig. 2, we show a magnified portion of
the high field 44.5 GHz data (highest $Q$ mode); SdH oscillations
are clearly visible. A fourier transform gives rise to a single
peak (main part of Fig. 2) at a frequency of 2975 T, in good
agreement with the $\alpha$-frequency reported previously
\cite{Mack1}. Thus, we can be extremely confident that we are well
coupled to the sample within the cavity. Weaker SdH oscillations
(due to lower $Q$-values) were discernible in all but the highest
frequency data. The absence of $\beta$ and $\gamma$ frequencies in
the SdH spectra is due to the relatively high temperature (1.4 K)
of these measurements.

Fig. 3 shows a compilation of the absorption peak-field positions
plotted against frequency for the four TE01$n$ cavity modes used
in this study. The exact peak-field positions were obtained from
fits to 1.4 K data. All of the points fall nicely on one of
several straight lines which pass through the origin, as expected
for cyclotron-like resonances, {\em i.e.} the resonance fields
(B$_{res} = m_c \omega / e$) scale linearly with frequency.

In order to interpret the data, it is first necessary to determine
the mechanism responsible for the apparent resonant dissipation
within the cavity. Due to the highly anisotropic conductivity in
Sr$_2$RuO$_4$, one can expect the inter-layer currents to
penetrate considerably further into the sample than the in-plane
currents, as illustrated in the inset to Fig. 1. At 1.5 K and 50
GHz, we estimate a penetration depth for {\bf{$ab$}}-plane
currents of $\delta_{ab}$ $\approx$ 0.2 $\mu$m; the corresponding
inter-layer penetration depth ($\delta_{c}$) is expected to be 30
to 40 ($\sim [\sigma_{ab}/\sigma_{c}]^{1/2}$) times greater. In
this so-called "skin depth" regime, when currents only flow close
to the sample surface, dissipation per unit area is governed by
the surface resistance R$_s$ $( \propto {\mathop{\rm Re}\nolimits}
\{ \sqrt {1/\hat \sigma } \} )$ \cite{dress}. Assuming a
quasi-static approximation, it is easy to see that for screening
of the AC fields ({\bf\~ H}$_1$) to occur within the bulk of the
sample, in-plane currents must flow predominantly over the large
flat surfaces, while inter-layer currents will flow at the sample
edges (see inset to Fig. 1). When we take into account a
geometrical factor ($\sim$ 20 for the sample used in this study),
corresponding to a ratio of the areas of the sample faces to its
edges, we find that $\sigma_{ab}$ and $\sigma_c$ contribute
more-or-less equally to dissipation within the cavity. This
contrasts the situation in layered organic conductors, where the
inter-layer conductivity tends to dominate such a measurement.

Next, we consider the symmetries of the three FSs in
Sr$_2$RuO$_4,$ and the possible resonance modes in $\sigma_{ab}$
and $\sigma_c$ resulting from the application of a DC field {\bf
B}$_o$//{\em c}. Three fundamental cyclotron modes should dominate
$\sigma_{ab}$, with cyclotron frequencies given by $\omega_{ci} =
e${\bf B}$_o$/$m_{ci}$, where the subscript {\em i} refers to the
band, {\em i.e.} $\alpha$, $\beta$ or $\gamma$ \cite{Mack1}.
Depending on the cross sectional shape of each FS section,
$\sigma_{ab}$ may additionally contain odd ($N =3,5, etc.$)
harmonics of the three fundamental CR frequencies \cite{Blundell};
these harmonics would occur at 1/{\em N} of the field of the
fundamental. Resonances in $\sigma_c$ are governed by the FS
warpings, which have recently been measured by Bergemann {\em et
al.} \cite{berg}. The strongest contributions to these warpings
have cylindrical symmetry and do not, therefore, affect motion in
the {\em c} direction. Nevertheless, weaker two-fold ($\alpha$)
and four-fold ($\alpha, \beta$ and $\gamma$) symmetric components
do exist \cite{berg}. Consequently, $\sigma_c$, may display weak
resonances at even {\em N} harmonics of the fundamental $\alpha$
CR frequency, and $N=4p$ ($p=$integer) harmonics of the
fundamental $\beta$ and $\gamma$ CR frequencies. The fundamental
frequencies themselves should not be observable in $\sigma_c$
\cite{berg,Hill1}.

We can now attempt to account for all of the observed resonances
in terms of the three bands ($\alpha, \beta$ and $\gamma$) in
Sr$_2$RuO$_4$. The inset to Fig. 3 shows an enlarged view of the
low field (high 1/{\bf B}) portion of the $f$ = 76.4 GHz data
plotted versus $\omega/\omega_c$, where $\omega$ = 2$\pi f$ and
$\omega_c$ is the cyclotron frequency (e{\bf B}/$m_c$). The
resonances labeled $N=1$ and $N=2$ correspond respectively to the
up ($\triangle$) and down ($\nabla$) triangle data points in the
main part of the figure. It is apparent that the $N=1$ peak is
actually the first in a harmonic series with resonances visible up
to $N=4$. Thus, we attribute all of these peaks to a single FS.
Further confirmation of this can be found from fits to the data in
the main part of Fig. 3 where the slopes obtained for the up and
down triangle data points differ by a factor of exactly 2
($\pm0.05$).

The remaining resonances are not related harmonically, either to
each other, or to the data in the inset to Fig. 3. Thus, we
conclude that the CR response is attributable to three independent
carrier types $-$ in agreement with both the theoretically
\cite{LDA} and experimentally determined FS \cite{Mack3}. The most
likely scenario is that the three main resonance peaks (above 8
tesla in Fig. 1) correspond to the three fundamental CR modes
associated with $\sigma_{ab}$. We assign the heaviest CR mass
$m_{c\gamma}$ = ($9.71\pm0.2$)$m_e$ to the $\gamma$-FS which, to
our knowledge, is the largest electronic effective mass thus far
detected by magnetic resonance. The CR mass for the $\beta$-FS is
$m_{c\beta}$ = ($5.81\pm0.05$)$m_e$, while we attribute the
harmonic series of resonances to the $\alpha$-FS with a CR mass
$m_{c\alpha}$ = ($4.33\pm0.05$)$m_e$. These values were deduced
from the slopes of the lines through the data in Fig. 3, thereby
avoiding any possible (systematic) errors associated with the
Lorentzian fits. Interpretation of the main absorption peaks as
$\sigma_c$ resonances would imply CR masses four times greater
than those deduced above, thus, effectively ruling out this
scenario. At this stage, it is not possible to ascertain whether
the $\alpha$ harmonics correspond to $\sigma_{ab}$ or $\sigma_{c}$
resonances; angle dependent studies should be able to resolve this
issue \cite{Hill1,Blundell}. The strong harmonic content of the
$\alpha$ CR is not unexpected given its approximately diamond
shaped cross section \cite{LDA}.



A comparison between the cyclotron masses deduced in this study
and the thermodynamic masses ($m_\alpha^*$= 3.4$m_e$, $m_\beta^*$=
7.5$m_e$ and $m_\gamma^*$= 14.6$m_e$) measured by Mackenzie {\em
et al}. [8], reveal clear enhancements of $m^*$ over $m_c$ for the
electron-like FSs ($\beta$ and $\gamma$). This does not appear to
be the case for the hole-like ($\alpha$-) FS. However, recent
theory by Kanki {\em et al.} \cite{kanki} has shown that if the
translational invariance of a FL is broken ({\em e.g.} due to the
lattice), $m_c$ may exceed $m^*$ under some circumstances.

According to FL theory, the ratio between $m^*$ and the so-called
dynamical mass which, in this case, we are assuming corresponds to
the cyclotron mass $m_c$, is given by $m^*$/$m_c$ = 1 + $F_1^s$/3,
where $F_1^s$ is the dimensionless $l = 1$ FL parameter
\cite{Quader}. In turn, $F_1^s$ should be proportional to the
thermodynamic density of states at $E_F$ which, for a
two-dimensional system, is proportional to the thermodynamic
effective mass at $E_F$, $i.e$. $F_1^s \propto m^*$. In fact, the
T-linear coefficient of the specific heat is in excellent
agreement with that estimated from $m^*$ on the assumption of the
FS two-dimensionality \cite{Mack3}. Consequently, one expects the
mass enhancements [($m^*$/$m_c$)$-$1] to scale with the
thermodynamic masses $m^*$. Interestingly, a comparison between
our results and the dHvA masses reported by Mackenzie {\em et al}.
\cite{Mack3}, reveal that the mass enhancements for the $\beta$
and $\gamma$ FSs scale most closely with the cyclotron masses
$m_{c\beta}$ and $m_{c\gamma}$, $i.e.$
[($m_\gamma^*$/$m_{c\gamma}$)$-$1]/[($m_\beta^*$/$m_{c\beta}$)$-$1]
= 1.73 and $m_{c\gamma}$/$m_{c\beta}$ = 1.67. Nevertheless, the
thermodynamic mass ratio ($m_\gamma^*$/$m_\beta^*$ = 1.95) is not
too far off from 1.73 either. Indeed, if one uses the $\beta$ dHvA
mass ($7.2m_e$) from ref. \cite{Yoshida} and the $\gamma$ dHvA
mass from ref. \cite{Mack3}, the mass enhancement ratio and the
thermodynamic mass ratio both come out close to 2. All said and
done, the very fact that the $\beta$ and $\gamma$ mass
enhancements seem to scale with the effective masses lends strong
support to the assertion that Sr$_2$RuO$_4$ is a correlated Fermi
liquid.

A word of caution is appropriate at this point. The underlying
theory used in the above analysis was developed for a single-band
isotropic FL. In Sr$_2$RuO$_4$, one might expect a strong coupling
between the various cyclotron modes corresponding to each electron
or hole band. In turn, these couplings might be expected to have a
pronounced effect on the measured cyclotron frequencies, over and
above the effects expected for a simple FL. With this in mind, it
is reasonable to expect the dominant electronic modes ($\beta$ and
$\gamma$) to have more of an influence on the hole mode ($\alpha$)
\cite{interact}, as opposed to the other way round, which might
explain why the $\alpha$-data do not follow the same trends seen
for the $\beta$ and $\gamma$ FSs.


Finally, we consider the temperature dependence of the resonances,
which attenuate dramatically above 3 K. This is not expected for a
single non-interacting parabolic band, unless the quasiparticle
lifetime $\tau$ depends strongly on temperature. One explanation,
therefore, is that the T$^2$ dependence of 1/$\tau$ is responsible
for this attenuation \cite{Maeno3}. However, we cannot rule out
the possibility that above $\sim$3 K, when $k_B$T exceeds
$\hbar\omega_c$, the CR lines broaden due to non-parabolicity.
Interactions would then further attenuate the resonances due to
the broadened distribution of cyclotron frequencies.


In summary, we have measured the cyclotron masses corresponding to
the $\alpha$, $\beta$ and $\gamma$ FSs in Sr$_2$RuO$_4$.
Comparisons between these values and those deduced from dHvA
studies reveal considerable enhancements of the thermodynamic
effective masses which we attribute to electron-electron
interactions. Qualitatively, our findings are in good agreement
with FL theory.


We are indebted to Christoph Bergemann, Andrew Mackenzie and
Stephen Julian for important discussions. This work was supported
by the Petroleum Research Fund (33727-G3) and the Office of Naval
Research (N00014-98-1-0538). Work carried out at the NHMFL was
supported by a cooperative agreement between the State of Florida
and the NSF under DMR-95-27035.


\end{multicols}
\clearpage

\begin{figure}
\centerline{\epsfig{figure=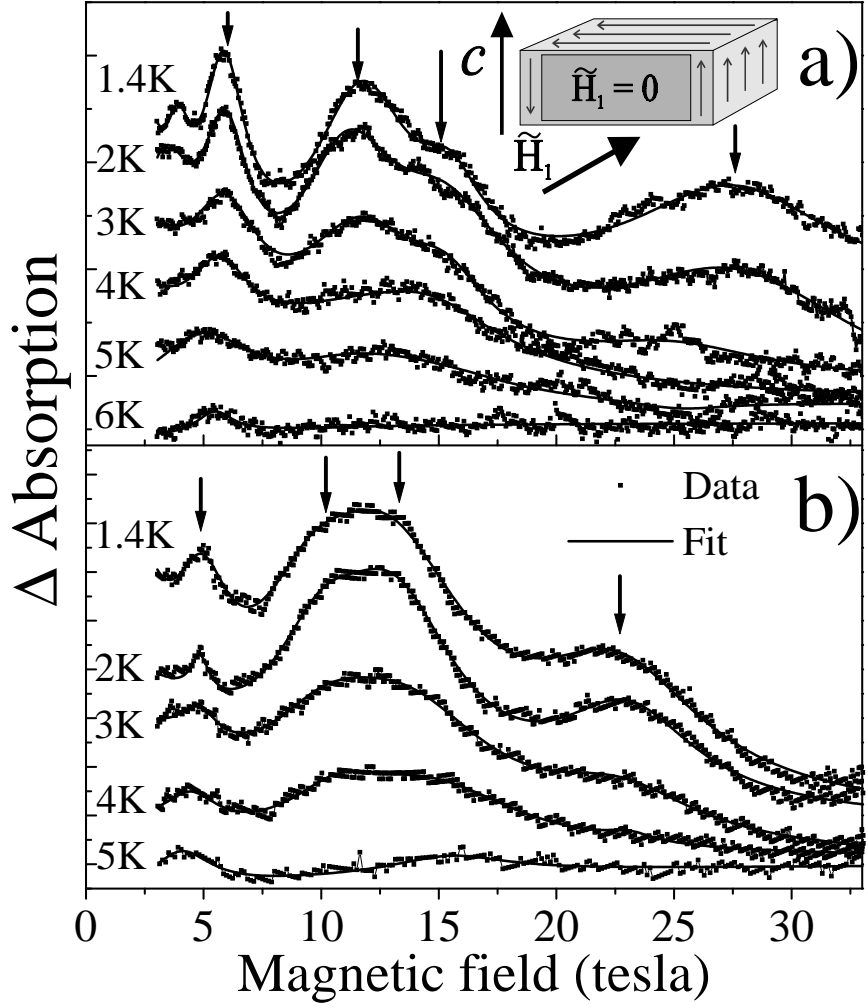,width=120mm}}
\bigskip
\caption{Temperature and magnetic field dependence of the changes
in absorption within the cavity (after a background subtraction)
at a) 76.4 GHz, and b) 64 GHz; the data have been offset for the
sake of clarity. See text for an explanation of the fitting
procedure. Inset depicts the predicted AC currents at the sample surface} 
\end{figure}

\clearpage

\begin{figure}
\centerline{\epsfig{figure=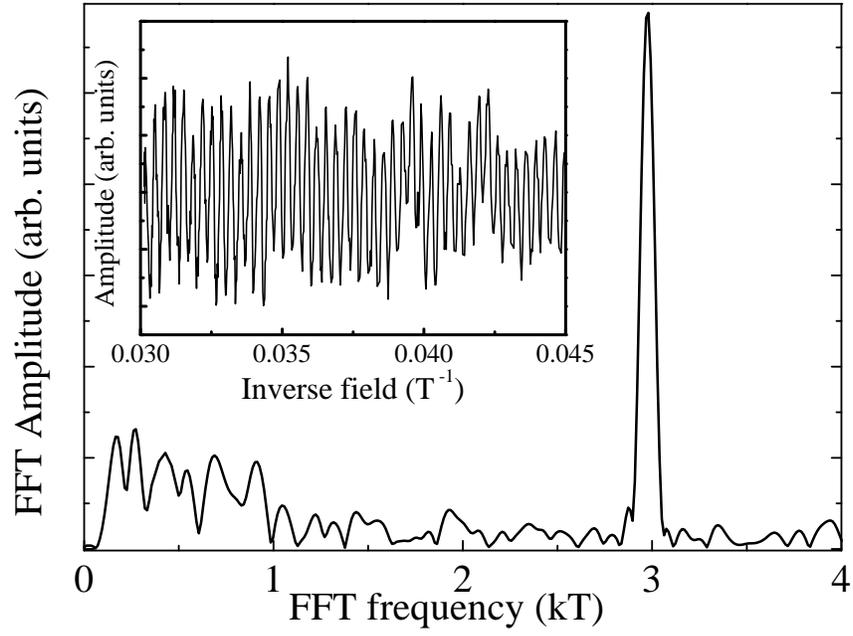,width=120mm}}
\bigskip
\caption{Fast fourier transform (FFT) of the high-field cavity
response (inset) at 44.5 GHz and at 1.4 K. SdH oscillations are
visible in the inset which are periodic in 1/{\bf B}. These
oscillations correspond to the $\alpha$-frequency observed in ref.
[3].} 
\end{figure}

\clearpage

\begin{figure}
\centerline{\epsfig{figure=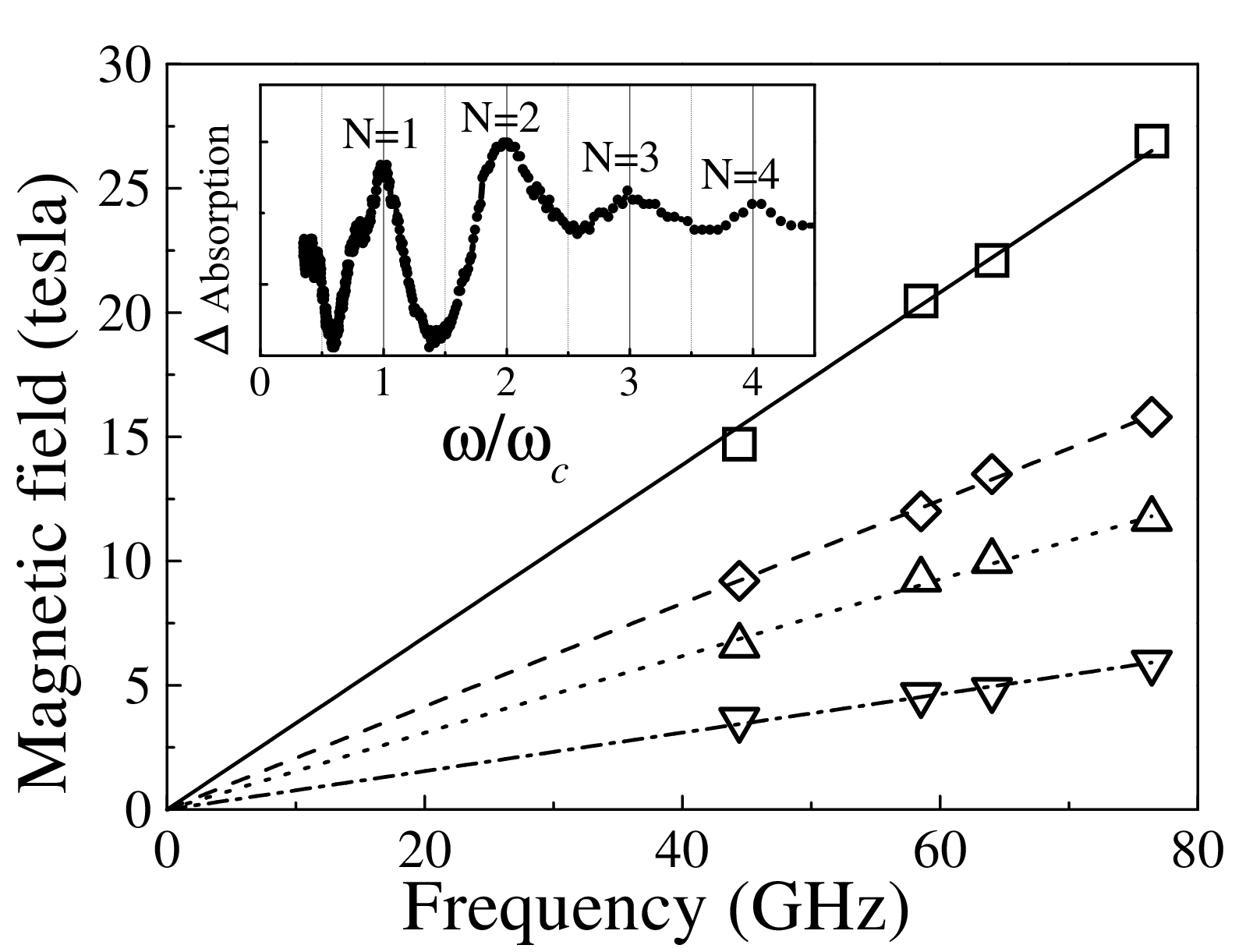,width=120mm}}
\bigskip
\caption{Plots of the CR peak-field positions versus frequency at
1.4 K. We assign the resonances as follows: $\Box - \gamma$,
$\diamondsuit - \beta$, $\triangle - \alpha$ 1st harmonic, and
$\nabla - \alpha$ 2nd harmonic. The inset shows the
$\alpha$-series of CR harmonics at 76.4 GHz, plotted versus
inverse field.} 
\end{figure}








\begin{references}
\bibitem{interact}
One might expect the strongest interaction to occur between $\alpha$ and $\beta$ which share the same
dominant $d_{yz}$ and $d_{zx}$ character.
\end{references}
\end{document}